\documentclass{ws-procs975x65}%
\usepackage{color}
\usepackage{amsmath}
\usepackage{amsfonts}
\usepackage{verbatim}
\usepackage{amssymb}
\usepackage{graphicx}
\usepackage{epstopdf}
\usepackage{mathrsfs}
\usepackage{epsfig}%
\providecommand{\U}[1]{\protect\rule{.1in}{.1in}}
\begin{document}

\title{Aspects of Stability of Hairy Black Holes}
\author{Andr\'{e}s Anabal\'{o}n$^{(1)}$ and Ji\v{r}\'{\i} Bi\v{c}\'{a}k$^{(2)}$\\\textit{$^{(1)}$ Universidad Adolfo Ib\'{a}\~{n}ez, Vi\~{n}a del Mar, Chile}\\\textit{$^{(2)}$ Institute of Theoretical Physics, Charles University, Prague,
Czech-Republic}}

\begin{abstract}
We analyze spherical and odd-parity linear perturbations of hairy black holes
with a minimally coupled scalar field.

\end{abstract}

%\maketitle

\section{Spherical Modes}

In order to have an asymptotically flat hairy black hole with a
minimally coupled scalar field a necessary condition is a scalar
field potential with a negative region. For a recent review see, for
example. \cite{Anabalon:2012dw}

As is well-known the stability problem can be mapped to the analysis
of the spectrum of a Schr\"{o}dinger operator, which appears in the
master equation for perturbations. An everywhere positive spectrum
implies there are no modes which exponentially grow in time. Using
the method of Ref.~\refcite{Bron1} we study the equations of motion
for the radial perturbations of the form
\begin{equation}
-{\frac{d^{2}u}{d\rho^{2}}}+V_{\mathrm{eff}}u=E^{2}u\qquad\hbox{with}\qquad
\delta\phi\propto e^{\mathrm{i}Et}u(\rho)\,, \label{schroe}%
\end{equation}
where $V_{\mathrm{eff}}(\rho)$ (explicitly given below) is an effective
potential in which the modes $u(\rho)$ propagate and $\rho$ is a
\textquotedblleft tortoise" radial coordinate sending the horizon at minus
infinity; it always exhibits a negative region. A sufficient condition for the
existence of bound states with negative $E^{2}$ (for bounded $V_{\mathrm{eff}%
}$ that fall-off faster than $\left\vert \rho\right\vert ^{-2}$)\ is the Simon
criteria, which states that if $S\equiv\int_{-\infty}^{+\infty}V_{\mathrm{eff}%
}\,d\rho$ is negative there will always be at least one bound state
with negative $E^{2}$; hence, we only study positive Simon
integrals. Using \textquotedblleft shooting" techniques to solve
(\ref{schroe}), we do indeed find unstable modes in
Ref.~\refcite{Anabalon:2012dw}. However, there is only a finite
number of unstable modes and, moreover, their characteristic time of
growth can be made arbitrarily large for certain values of the black
holes parameters, as is the case if the size of the black hole is
small enough.

We are interested in studying the linearized dynamics around a background
solution. Hence, starting from the metric in the form
\begin{equation}
ds^{2}=-\left[  A(r)+\epsilon A_{1}(r,t)\right]  dt^{2}+\left[  B(r)+\epsilon
B_{1}(r,t)\right]  dr^{2}+C(r)d\Omega^{2}\text{ ,}\label{coor}%
\end{equation}
where $r$ is a general radial coordinate, not necessarily $\rho$, and the
scalar field is assumed in the form $\phi=\phi_{0}(r)+\epsilon\phi_{1}(r,t)$.
We expand the scalar field potential as $V(\phi)=V_{0}+\epsilon V_{1}\phi
_{1}(r,t)$ where $V_{0}=V(\phi_{0})$ $,V_{n}=\left.  \frac{d^{n}V}{d\phi^{n}%
}\right\vert _{\phi=\phi_{0}}$. As a consequence of spherical symmetry all the
dynamics is driven by the scalar field. Indeed, it is possible to write the
metric perturbations in terms of the $\phi_{1}(r,t)$ by using the Einstein
field equations. We introduce the master variable
\begin{equation}
\psi(\rho,t)=\phi_{1}(r,t)C(r)^{1/2}\text{ ,}%
\end{equation}
where $\rho~$is the tortoise coordinate%
\begin{equation}
d\rho=\left(  \frac{B}{A}\right)  ^{1/2}dr\text{ .}\label{tor}%
\end{equation}
The master equation is
\begin{equation}
-\partial_{\rho}^{2}\psi+V_{\mathrm{eff}}\psi=-\partial_{t}^{2}\psi\text{
,}\label{Schr}%
\end{equation}
with the effective potential%
\begin{equation}
\frac{V_{\mathrm{eff}}}{A}=4\kappa C\left[  \left(  \kappa V_{0}C-1\right)
\left(  \frac{d\phi_{0}}{dC}\right)  ^{2}+V_{1}\left(  \frac{d\phi_{0}}%
{dC}\right)  \right]  -\kappa V_{0}+V_{2}+\frac{1}{C}-\frac{1}{4B}\left(
\frac{C^{\prime}}{C}\right)  ^{2}\text{ }.\label{pot}%
\end{equation}

If $\rho$ takes its values in the whole real line and $V_{\mathrm{eff}}$ is
non-negative, the operator (\ref{Schr}) is essentially self-adjoint and its
spectrum is positive which implies that the background is mode stable under
spherically symmetric perturbations.

As mentioned above, using the shooting method, we found the
asymptotically flat black holes unstable with respect to spherically
symmetric perturbations. However, if their mass $M$ and size $r_{+}$
are small compared with the coupling constant appearing in the
scalar field potential, the time for the instability to develop is
long compared to the scale set by $M$. More details are given in
Ref.~\refcite{Anabalon:2013baa}, see also.\cite{Kleihaus:2013tba}

For asymptotically anti-de Sitter boundary conditions, the tortoise
coordinate takes its values in the half real line $\rho\in\left]
-\infty,0\right]  $. If $V_{\mathrm{eff}}$ is non-negative, the
spectrum can still contain a negative eigenvalue which depends on
the details of the theory and the boundary conditions of the scalar
field. This was recently discussed in more detail by one of us
in.\cite{Anabalon:2015vda}

\section{Odd-Parity Modes}

In the second part we analyze the stability of hairy black holes
under odd-parity perturbations following our work in
Ref.~\refcite{Anabalon:2014lea}. In contrast to the radial
perturbations in asymptotically flat spacetimes we show that
independently of the scalar field potential and of specific
asymptotic properties of spacetime (asymptotically flat, de Sitter
or anti-de Sitter), any static, spherically symmetric or planar,
black hole solution of the Einstein theory minimally coupled to a
real scalar field with a general potential is mode stable under
linear odd-parity perturbations. We analyze their odd-parity
perturbations following the general treatment of the
\textquotedblleft axial\textquotedblright\ perturbations of
spherically symmetric spacetime which are \textit{not} necessarily
vacuum, by
Chandrasekhar. The perturbed metric reads%
\begin{equation}
ds^{2}=-Adt^{2}+Bdr^{2}+C\left[  \frac{dz^{2}}{\left(  1-kz^{2}\right)
}+\left(  1-kz^{2}\right)  \left(  d\varphi+k_{1}dt+k_{2}dr+k_{3}dz\right)
^{2}\right]  \ ,
\end{equation}
where $k_{1}$, $k_{2}$ and $k_{3}$ are functions of ($t,r,z$), $A(r),$ $B(r)$
and $C(r)$ are the metric functions parameterizing the most general static
background solution of a scalar-tensor theory. For asymptotically locally AdS
solutions, $k=\pm1\,\ $or $0$. Asymptotically flat or de Sitter solutions have
$k=1$. The scalar field is taken to be of the form $\phi=\phi_{0}%
(r)+\epsilon\Phi\left(  t,r,z\right)  $, where $\phi_{0}$ is the background
field. The metric perturbations $\left(  k_{1},k_{2},k_{3}\right)  $ are all
taken to be first order in $\epsilon$. Since any surface of constant $(t,r)$
is of constant curvature, we consider only axisymmetric perturbations, without
any loss of generality. The Einstein field equations are truncated at first
order in $\epsilon$. This yields the vanishing of $\Phi$. Introducing the
variable $Q=CA^{1/2}B^{-1/2}(1-kz^{2})^{2}\left(  \partial_{z}k_{2}%
-\partial_{r}k_{3}\right)  $ and assuming $Q=q(r,t)D(z)$, we find that a
combination of the field equations yield
\begin{align}
\frac{C^{2}}{\sqrt{AB}}\frac{\partial}{\partial r}\left[  \frac{A^{1/2}%
}{CB^{1/2}}\frac{\partial q}{\partial r}\right]  -\lambda q  &  =\frac{C}%
{A}\partial_{t}^{2}q\text{ },\label{eq}\\
\left(  1-kz^{2}\right)  ^{2}\frac{\partial}{\partial z}\left[  \frac
{1}{\left(  1-kz^{2}\right)  }\frac{\partial D}{\partial z}\right]   &
=-\lambda D\text{ }, \label{gegenbauer}%
\end{align}
where $\lambda$ is a separation constant. Let us put $k=1$ and set
$z=\cos\theta$ in equation (\ref{gegenbauer}); then $C_{l+2}^{-3/2}%
(\theta)=D(z)$ is the Gegenbauer polynomial with $\lambda=\left(  l-1\right)
\left(  l+2\right)  $, $l\geq1$ holds. The master variable in this case is
$\Psi(\rho,t)=q(r,t)C^{-1/2}$ where $\frac{\partial}{\partial r}=\frac
{B^{1/2}}{A^{1/2}}\frac{\partial}{\partial\rho}$. Fourier decomposing the
master variable, $\Psi(\rho)=\int\Psi_{\omega}e^{i\omega t}dt$, yields the
master equation
\begin{equation}
\mathcal{H}\Psi_{\omega}\equiv-\frac{d^{2}\Psi_{\omega}}{d\rho^{2}}+\left(
\lambda\frac{A}{C}+\frac{3}{4C^{2}}\left(  \frac{dC}{d\rho}\right)  ^{2}%
-\frac{1}{2C}\frac{d^{2}C}{d\rho^{2}}\right)  \Psi_{\omega}=\omega^{2}%
\Psi_{\omega}\text{ }. \label{RW}%
\end{equation}
The scalar field perturbation vanishes, however equation (\ref{RW}) depends on
the background scalar field through its influence on the background metric (in
vacuum, the equation (\ref{RW}) becomes the Regge-Wheeler equation). The
operator $\mathcal{H}$ is not manifestly positive, but its spectrum is
positively defined as has been shown by finding a suitable $S-$deformation.

\section{Slowly Rotating Hairy Black Holes}

Consider a stationary perturbations with $k_{2}=k_{3}=0$ and $k_{1}=\omega
(r)$. In this case we find%

\begin{equation}
\omega=-c_{1}\int\frac{\sqrt{AB}}{C^{2}}dr+c_{2}\text{ }, \label{omega}%
\end{equation}
where $c_{1}$ and $c_{2}$ are two integration constants. Let us first consider
the case of the Schwarzschild black hole. We have $\sqrt{AB}=1$ and $C=r^{2},$
so it follows that%
\begin{equation}
\omega=\frac{c_{1}}{3r^{3}}+c_{2}\text{ }.
\end{equation}
Hence, choosing $c_{2}=0$ and $c_{1}=3Ma,$ we find the slowly rotating Kerr
black hole.

Now let us consider the hairy black hole family reviewed in
Ref.~\refcite{Anabalon:2012dw}. In analogy with the Kerr solution,
the slowly rotating
hairy black hole is a deformation of the static one plus $g_{t\varphi}%
=\omega_{\nu}(1-z^{2})C(r)$, $C(r)$ is the areal function. The metric
component $g_{t\varphi}$ determines the frame dragging potential. We find that%
\begin{equation}
\omega_{\nu}=\bar{c}_{1}\left(  \frac{r^{2-\nu}}{\nu^{2}(\nu^{2}-4)}(\left(
\nu-2\right)  r^{2\nu}+\left(  4-\nu^{2}\right)  r^{\nu}-2-\nu)+\frac{1}%
{\nu^{2}-4}\right)  \text{ .}%
\end{equation}
To measure the deviation of the dragging effects from those from the
slowly rotating Kerr solution we plot the ratio
$\omega/\omega_{\nu=1}$ versus the square root of the areal function
$\sqrt{C(r)}$. In Figure $1$ it can be seen that there is a smooth
departure from the Kerr frame dragging as both coincide when $\nu$
approaches $1$ or asymptotically for large $\sqrt{C(r)}$. It should
be noticed that the departure from Kerr dragging can be important
and that the horizon can be located at any point in the graph.
Indeed, the location of the horizon is defined by the equation
$A(r_{+})=0$, which has a solution for any $r_{+}$ by adjusting the
value of the other parameters in the metric, see.
\cite{Anabalon:2014lea}

\bigskip We acknowledge the Grant No. 14-37086G of the Czech Science Foundation (Albert Einstein Centre).

\begin{figure}[ptb]
\centering
\epsfig{file=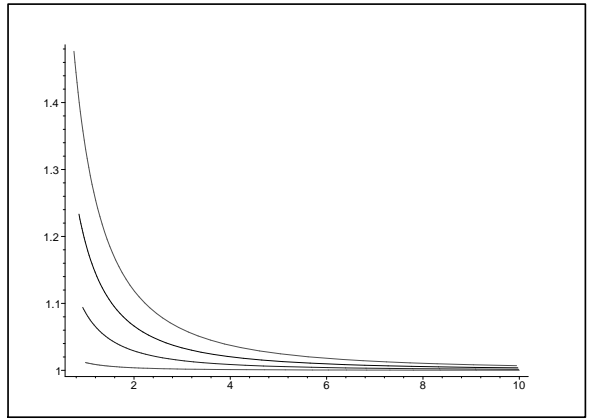,width=5.5cm,angle=270}\caption{The ratio
$\omega/\omega_{\nu=1}$ versus the square root of the areal function,
$\sqrt{C(r)}$, for different values of $\nu$. The plots are for $\nu
=1.2,\nu=2.1,\nu=3$ and $\nu=4$ (from down up).}%
\end{figure}

\end{document}